\documentclass[preprint,aps]{revtex4}
\newcommand{\be}{\begin{equation}}
\newcommand{\ee}{\end{equation}}
\newcommand{\bea}{\begin{eqnarray}}
\newcommand{\eea}{\end{eqnarray}}
\newcommand{\ba}{\begin{array}}
\newcommand{\ea}{\end{array}}

\begin{document}
\title{Weak vector and axial-vector form factors in the chiral constituent quark model with configuration mixing}
\author{Neetika Sharma$^a$, Harleen Dahiya$^a$, P.K. Chatley$^a$, Manmohan Gupta$^b$}
\address{$^a$Department of Physics, Dr. B.R. Ambedkar National
Institute of Technology, Jalandhar, 144011, India
\\ $^b$ Department of Physics, Centre of Advanced Study in Physics, Panjab
University, Chandigarh 160014, India}

\begin{abstract}
The effects of SU(3) symmetry breaking and configuration mixing
have been investigated for the weak vector and axial-vector form
factors in the chiral constituent quark model ($\chi$CQM) for the
strangeness changing as well as strangeness conserving
semi-leptonic octet baryon decays in the nonperturbative regime.
The results are in good agreement with existing experimental data
and also show improvement over other phenomenological models.
\end{abstract}
\maketitle


The measurements in the deep inelastic scattering (DIS)
experiments \cite{emc} indicate that the valence quarks of the
proton carry only about 30\% of its spin and also establishes the
asymmetry of the quark distribution functions \cite{e866}.
Further, these  measurements relate the spin dependent
Gamow-Teller matrix elements to the weak vector and axial-vector
form factors ($f_{i=1,2,3}(Q^2$) and $g_{i=1,2,3}(Q^2$)) of the
semi-leptonic baryon decays \cite{okun}. These form factors
provide vital information on the interplay between the weak
interactions (low-$Q^2$) and strong interactions (large-$Q^2$) and
are an important set of parameters for investigating in detail the
dynamics of the hadrons particularly at low energies.

The baryons are usually assigned to a SU(3)-flavor octet to deduce
 the spin densities and their relation with the weak matrix
 elements of the semi-leptonic decays \cite{okun}.
The data to study the form factors was earlier analyzed under the
assumptions of exact SU(3) symmetry \cite{cern-WA2}. However, the
experiments performed in the late eighties were more precise and
the assumption of SU(3) symmetry could no longer provide a
reliable explanation of the form factors indicating that SU(3)
symmetry breaking effects are important. This was first observed
for the $\Sigma^- \rightarrow n e^{-} \bar{\nu_e}$ decay with the
measurement of $ |(g_1 - 0.133g_2)/f_1|$ = 0.327 $\pm 0.007 \pm
0.019$   giving $\frac{g_1}{f_1}=-0.20 \pm 0.08$ and
$\frac{g_2}{f_1}=-0.56 \pm 0.37$ \cite{syh}. These values were
quite different from the results obtained assuming SU(3) symmetry.
The importance of SU(3) symmetry breaking has been further
strengthened from the $\frac{g_1}{f_1}$ ratio of the $\Xi^0
\rightarrow \Sigma^+ e^- \bar {\nu_e}$ decay measured by KTeV
(Fermilab E799) experiment \cite{ktev} giving
$1.32^{+0.21}_{-0.17} \pm 0.05$, with the assumption of SU(3)
symmetry and $1.17 \pm 0.28\pm 0.05$, in the limit of SU(3)
breaking. Recently, this decay has been studied by NA48/1
Collaboration \cite{batley} giving $\frac{g_1}{f_1}$ = 1.20 $\pm$
0.05 which is more in agreement with the results of the KTeV
experiment in the limit of SU(3) symmetry breaking.

Theoretically, the question of
SU(3) symmetry breaking  has  been investigated  by several
authors using various phenomenological models. Calculations have been
carried out for the weak form factors in the Cabibbo model
\cite{cabibbo2003} assuming exact SU(3) symmetry, chiral
quark-soliton model (CQSM) \cite{kim,tim}, relativistic
 constituent quark model(RCQM) \cite{felix}, Yamanishi's
 model using mass splitting interactions (MSI)
\cite{yama}, $1/N_c$ expansion of QCD \cite{ruben,matuea}, chiral
perturbation theory (ChPT) \cite{lac,ruben1}, lattice QCD
\cite{gaud}, covariant chiral quark approach (CCQ) \cite{cc} etc..
The predictions of these models are however not in agreement with
each other in terms of the magnitude as well as the sign of these
form factors. Therefore, it would be interesting to examine the
spin structure and the weak form factors of the baryons at low
energy, thereby giving vital clues to the nonperturbative effects
of QCD.

It has been shown recently that the chiral constituent quark model
($\chi$CQM) \cite{manohar} has been successful in explaining
various general features of the quark flavor and spin distribution
functions \cite{cheng} and baryon magnetic moments \cite{cheng}.
Also, it has been shown that configuration mixing generated by
spin-spin forces \cite{{dgg}}, known to be compatible with the
$\chi$CQM (henceforth to be referred as $\chi$CQM$_{{\rm
config}}$), improves the predictions of $\chi$CQM regarding the
spin polarization functions \cite{hd} and is able to give an
excellent fit \cite{hdorbit} to the baryon magnetic moments. The
purpose of the present work is to carry out a detailed analysis of
the weak vector and axial-vector form factors at low energies for
the semi-leptonic decays of baryons within the framework of
$\chi$CQM$_{{\rm config}}$. In particular, we would like to
calculated the individual vector and axial-vector form factors
($f_{i=1,2,3}(Q^2$ and $g_{i=1,2,3}(Q^2)$) as well as the ratios
of these form factors for both the strangeness changing ($\Delta
S=1$) as well as strangeness conserving ($\Delta S=0$) decays.
Further, it would be interesting to understand in detail the role
of SU(3) symmetry breaking  in the weak axial-vector form factors.

The effective Lagrangian in the $\chi$CQM formalism describes the
interaction between quarks and a nonet of Goldstone bosons (GBs)
where the fluctuation process is $q^{\pm} \rightarrow {\rm GB} +
q^{' \mp} \rightarrow  (q \bar q^{'}) +q^{'\mp}$ \cite{cheng,hd}.
The GB field is written as \bea
 \Phi = \left( \ba{ccc} \frac{\pi^0}{\sqrt 2}
+\beta\frac{\eta}{\sqrt 6}+\zeta\frac{\eta^{'}}{\sqrt 3} & \pi^+
  & \alpha K^+   \\
\pi^- & -\frac{\pi^0}{\sqrt 2} +\beta \frac{\eta}{\sqrt 6}
+\zeta\frac{\eta^{'}}{\sqrt 3}  &  \alpha K^0  \\
 \alpha K^-  &  \alpha \bar{K}^0  &  -\beta \frac{2\eta}{\sqrt 6}
 +\zeta\frac{\eta^{'}}{\sqrt 3} \ea \right)\,. \eea The SU(3)$\times$U(1) symmetry breaking is introduced by
considering $m_s > m_{u,d}$ as well as by considering the masses
of GBs to be nondegenerate $(M_{K,\eta} > M_{\pi}$ and
$M_{\eta^{'}} > M_{K,\eta})$ \cite{cheng}.  The parameter
$a(=|g_8|^2$) denotes the probability of chiral fluctuation  $u(d)
\rightarrow d(u) + \pi^{+(-)}$, whereas $\alpha^2 a$, $\beta^2 a$
and $\zeta^2 a$ respectively denote the probabilities of
fluctuations $u(d) \rightarrow s + K^{-(0)}$, $u(d,s) \rightarrow
u(d,s) + \eta$, and $u(d,s) \rightarrow u(d,s) + \eta^{'}$.

Further, to make the transition from $\chi$CQM to $\chi$CQM$_{{\rm
config}}$, the nucleon wavefunction is modified because of the
configuration mixing generated by the chromodynamic spin-spin
forces \cite{dgg,hd} and the modified spin polarization functions
$\Delta q=q^+-q^-$ of different quark flavors can be taken from
Ref. \cite{hd}. It would be important to mention here that the
SU(3) symmetric calculations can easily be obtained by considering
$\alpha, \beta=1$ and $\zeta = -1$.

The matrix elements for the vector and axial-vector current in the
case of weak hadronic current $J{^\mu_h}$ for the semi-leptonic
hadronic decay process $B_i \rightarrow B_f +l +\bar {\nu_l}$ are
given as \cite{tommy,renton} \be \langle
B_f(p_f)|J{^\mu_V}|B_i(p_i)\rangle=\bar u_f(p_f) \left( f_1(Q^2)
\gamma^\mu - i\frac{f_2(Q^2)}{M_i+M_f} \sigma^{\mu\nu} q_\nu
+\frac{f_3(Q^2)}{M_i + M_f}q^\mu \right) u_i(p_i)\,, \label{jv}
\ee \be \langle B_f(p_f)|J{^\mu_A}|B_i(p_i)\rangle = \bar
u_f(p_f)\left ( g_1(Q^2)\gamma ^\mu \gamma^5 -i\frac{g_2(Q^2)}{M_i
+ M_f}\sigma^{\mu\nu} q_\nu \gamma^5 +\frac{g_3(Q^2)}{M_i +
M_f}q^\mu \gamma^5 \right) u_i(p_i)\,,\label{ja} \ee where  $M_i$
$(M_f)$ and $u_i(p_i)$ ($\bar u_f(p_f)$) are the masses and Dirac
spinors of the initial (final) baryon states, respectively. The
four momenta transfer is given as $Q^2 = -q^2$, where $ q \equiv
p_i - p_f$. The functions $f_i(Q^2)$ and $g_i(Q^2)$ ($i=1,2,3$)
are the dimensionless vector and axial-vector form factors.

Since we are interested to calculate the form factors at low
$Q^2$, in this context the generalized Sachs form factors at $Q^2
\approx 0$ can be introduced following Ref. \cite{tommy} and the
vector as well as axial-vector functions can be expressed in terms
of these generalized form factors. Similarly, the generalized
Sachs form factors at $Q^2 \approx 0 $  at the quark level can be
introduced following Ref. \cite{tommy}. In the nonrelativistic
limit, the current operators act additively on the three quarks in
the baryons, therefore, the Sachs form factors for the quark
currents can be used to obtain the corresponding Sachs form
factors for the baryons.  The vector and axial-vector form factors
can respectively  be expressed as \be f_1 = f_1(0)\,,~~~~~ f_2 =
\left(\frac{\Sigma M}{\Sigma m }\frac{G_A}{G_V}-1\right)f_1(0)\,,
~~~~~f_3 = \frac{\Sigma M}{\Sigma m}\left(E \frac{G_A}{G_V}
-\epsilon\right)f_1(0)\,, \label{f1-f3} \ee \be g_1 =
g{_1}(0)\,,~~~ g_2 = \left(\frac{\Sigma M}{\Sigma m}\epsilon
-\frac{1}{2}(1+\frac{\Sigma M^2}{\Sigma m^2})E\right) g_1(0)\,,~~~
g_3 = \left( \frac{1}{2}(1-\frac{\Sigma M^2}{\Sigma m^2}
)+\frac{\Sigma M ^2}{\Sigma m^2}g^q_3\right)g_1(0)\,,
\label{g1-g3} \ee where $\Sigma M=M_i+M_{f}$, $\Delta
M=M_i-M_{f}$, $\Sigma m=m_q+m_{q'}$, $\Delta m=m_q-m_{q'}$ and
$g_3^q$ is the induced pseudoscalar form factor at the quark
level. Only the linear part of symmetry breaking terms are being
calculated where the higher order terms involving $E \equiv
\frac{\Delta M}{\Sigma M}$ and $\epsilon\equiv \frac{\Delta
m}{\Sigma m} $ can be neglected. The baryon decays considered in
the present work are $n\rightarrow p$, $\Sigma^ \mp \rightarrow
\Lambda$, $\Sigma^-\rightarrow \Sigma^0$ and $ \Xi^- \rightarrow
\Xi^0$ corresponding to the strangeness conserving decays and
$\Sigma^- \rightarrow n$, $\Xi^- \rightarrow \Sigma^0$, $\Xi^-
\rightarrow \Lambda$, $\Lambda \rightarrow p$ and
$\Xi^0\rightarrow \Sigma^+$ corresponding to the strangeness
changing decays.

We now discuss the input parameters used in the calculations. To
begin with, we discuss the parameters involved in the calculation
of quark spin polarization functions. The $\chi$CQM$_{{\rm
config}}$ involves five parameters, four of these $a$, $a
\alpha^2$, $a \beta^2$, $a \zeta^2$ representing respectively the
probabilities of fluctuations to pions, $K$, $\eta$, $\eta^{'}$,
following the hierarchy $a > \alpha > \beta > \zeta$, while the
fifth representing the mixing angle. The mixing angle $\phi$ is
fixed from the consideration of neutron charge radius \cite{dgg},
whereas for the other parameters, we use the latest data
\cite{PDG}. In this context, it is found convenient to use $\Delta
u$, $\Delta_3$, asymmetries of the quark distribution functions
($\bar u-\bar d$ and $\bar u/\bar d$) as inputs with their latest
values given in Table I. Before carrying out the fit to the above
mentioned parameters, we determine their ranges by qualitative
arguments. To this end, the range of the symmetry breaking
parameter $a$, $\alpha$, $\beta$ and $\zeta$ are found to be $0.09
\lesssim a \lesssim 0.15$, $0.2 \lesssim \alpha \lesssim 0.5$,
$0.2\lesssim \beta \lesssim 0.7$ and $-0.65 \lesssim \zeta
\lesssim -0.08$ respectively.  After finding the ranges, we have
carried out a fine grained analysis using the above ranges as well
as considering $\alpha\approx \beta$ leading to $a=0.12$,
$\zeta=-0.15$, $\alpha=\beta=0.45$ as the best fit values. For the
$u$, $d$ and $s$ quarks, we have used their widely accepted values
in hadron spectroscopy \cite{cheng}, viz., $m_u$ = $m_d$ = 0.330
GeV, and $m_s$ = 3$m_u$/2 = 0.495 GeV. For evaluating the
contribution of GBs, we have used their on mass shell value in
accordance with several other similar calculations \cite{{mpi1}}.

In Table \ref{fsandgs}, we have given the individual values of
vector and axial-vector form factors in the $\chi$CQM$_{{\rm
config}}$ using the input values discussed earlier. Even though
there is no experimental data available for these form factors,
the individual values are important to compare our results with
other model calculations.  It can be clearly seen from the results
that the contributions of second class currents $f_3$ and $g_2$
are very small for the same isospin multiplets, for example,
$n\rightarrow p$, $\Sigma^-\rightarrow \Sigma^0$ and $ \Xi^-
\rightarrow  \Xi^0$. This is because of the small mass difference
between the initial and final decay particles. Also, for all other
decays, the second class currents are having a comparatively
smaller contribution than the other first class currents as
expected.

In Table \ref{g1f1}, we have presented the values of
$\frac{g_1}{f_1}$ at $Q^2$= 0 and compared our results with other
model calculations as well as the available experimental data. The
ratio of $g_1$ and $f_1$  is the non-singlet combination of the
quark spin polarizations given as $\Delta_3=\Delta u-\Delta d=
\frac{G_A}{G_V} =\frac{g_1(0)}{f_1(0)}$. We have also investigated
in detail the implications of SU(3) symmetry breaking and
presented the results, both with and without SU(3) symmetry
breaking. We are able to give a fairly good account for most of
the weak form factors (where the experimental data is available),
in line with the success of $\chi$CQM$_{{\rm config}}$ in
describing the spin dependent polarization functions. Our results,
in the case of ${\frac{G_A}{G_V}}^{\Sigma^- \rightarrow n}$,
${\frac{G_A}{G_V}}^{\Xi^- \rightarrow \Lambda}$,
${\frac{G_A}{G_V}}^{\Lambda \rightarrow p}$ and
${\frac{G_A}{G_V}}^{\Xi^0 \rightarrow \Sigma^+}$,  show a clear
improvement over the results of other calculations. It is also
interesting to consider the ratio
$\frac{(g_1/f_1)^{\Lambda\rightarrow p} }{(g_1/f_1)^{\Sigma^-
\rightarrow n}}$, which comes out to be $-2.34$ in our calculation
and is quite close to the experimental value $-2.11 \pm 0.15$
\cite{tim}.

In case of weak magnetism form factor ratio $\frac{f_2}{f_1}$,
experimental data is available only for two strangeness changing
decays. The results have been presented in Table \ref{f2f1}. In this
case also, the predictions of different models differ significantly
from each other. Our prediction for the $ \Xi^0 \rightarrow
\Sigma^+ $ decay matches well with the experiment. In the case of
$\Sigma^-\rightarrow n$ decay, it seems that our prediction for
the $\frac{f_2}{f_1}(=-1.81)$ is not in agreement with the
experimental value ($0.96\pm 0.15$) listed  in Ref. \cite{syh}.
However, it would be important to mention here that the above
 mentioned experimental value has been obtained with the assumption
of $g_2= 0$ or SU(3) symmetry. A better agreement can be
 found for $|(g_1 - 0.133g_2)/f_1|$ = $0.327 \pm 0.007 \pm
0.019$ where our prediction for this quantity is $0.31$, in fair
agreement with the data, which is clearly due to SU(3) breaking
effect. Pending further experimental data, we have predicted the
value of $\frac{f_2}{f_1}$ for all other baryon decays with and
without SU(3) symmetry.

To summarize, the chiral constituent quark model with
configuration mixing ($\chi$CQM$_{{\rm config}}$) and SU(3)
symmetry breaking is able to provide a fairly good description of
the weak vector and axial-vector form factors for the
semi-leptonic octet baryon decays. Our results are consistent with
the latest experimental measurements as well as with the lattice
QCD results and also show improvement over other phenomenological
models in some cases. A refinement in the case of the measurements
with the assumption of SU(3) symmetry breaking would have
important implications for the basic tenets of $\chi$CQM. In
conclusion, we would like to state that SU(3) symmetry breaking
and configuration mixing in the $\chi$CQM are the key in
understanding the hadron dynamics in the nonperturbative regime.

 {\bf ACKNOWLEDGMENTS}\\
H.D. would like to thank DST, Government of India, for financial
support.

\begin{table}
 {\footnotesize \begin{center}
\begin{tabular}{ccccccc} \hline
Decay & $ f_1$  & $f_2$ & $f_3$ & $g_1$ & $g_2$ & $g_3 $\\ \hline
$n\rightarrow p e^{-} \bar{\nu}$ & 1.00 & 2.612 & 0.003& 1.270
&$-$0.004& $-$232.9\\ $ \Sigma^- \rightarrow \Sigma^0 e^{-}
\bar{\nu} $ & 1.414 & 1.033 & 0.005 & 0.676 & $-$0.010 & $-$201.3
\\ $ \Sigma^-\rightarrow \Lambda e^{-} \bar{\nu}$&0 & 2.265 &
0.080 & 0.646 & $-$0.152 & $-$271.4\\ $  \Sigma ^+ \rightarrow
\Lambda e^{-} \bar{\nu}$ &0 & 2.257 & 0.072 & 0.646 & $-$0.136 &
$-$245.9\\ $ \Xi^- \rightarrow \Xi^0 e^{-} \bar{\nu} $ & $-$1.00 &
2.253 & 0.003 & 0.314 & $-$0.007 & 113.8\\ \hline
$\Sigma^-\rightarrow n e^{-} \bar{\nu}$ & $-$1.0 & 1.813 & 0.616 &
0.314 & 0.017 & $-$9.2\\ $\Xi^-\rightarrow \Sigma^0 e^{-}
\bar{\nu}$ & 0.707 & 2.029 & $-$0.291 & 0.898 & 0.310 & $-$29.1\\
$\Xi^-\rightarrow \Lambda e^{-} \bar{\nu}$ & 1.225 & $-$0.450 &
$-$0.658 & 0.262 &0.047 & $-$8.9 \\ $\Lambda \rightarrow p e^{-}
\bar{\nu}$& $-$1.225 & $-$1.037 & 0.415 & $-$0.909 & $-$0.170 &
20.7\\ $ \Xi^0 \rightarrow \Sigma^+ e^{-} \bar{\nu}$& 1.0 & 2.854
& $-$0.414 & 1.27 & 0.446 & $-$40.7\\ \hline
 \end{tabular}
\caption{Weak vector and axial-vector form factors for the
semi-leptonic octet baryon decays in the $\chi$CQM$_{{\rm
config}}$.} \label{fsandgs}
\end{center}}
\end{table}

\begin{table}
{\footnotesize \begin{center}
\begin{tabular}{cc cccc cccc}  \hline
Decay & Data  & RCQM   & CQSM   & MSI & ChPT & CCQ & $\chi$CQM &
$\chi$CQM$_{{\rm config}}$  & $\chi$CQM$_{{\rm config}}$ with \\
&\cite{PDG} & \cite{felix}&\cite{tim} & \cite{yama} &\cite{ruben1}
&\cite{cc} &
 \cite{tommy}& with SU(3) & SU(3) symmetry \\
 & &  & &  & & &
 & symmetry &  breaking \\      \hline
$\frac{G_A}{G_V}^{n\rightarrow p}$ & 1.2695 $\pm$ 0.0029  & 1.25 &
1.18& 5.3* $10^{-7}$&1.27 &1.27& 1.26& 0.95& 1.27\\
$\frac{G_A}{G_V}^{\Sigma^- \rightarrow \Sigma^0} $ & --  & 0.49&
0.46 & --& --&--& 0.5 & 0.39&0.48\\
$\frac{G_A}{G_V}^{\Sigma^-\rightarrow \Lambda}$  &
${\frac{f_1}{g_1}=0.01\pm 0.1 }$  & 0.74 & 0.73 & --& 0.62&0.62
&0.62& 0.45$^*$& 0.65$^*$\\ $\frac{G_A}{G_V}^{\Sigma ^+
\rightarrow \Lambda}$ &-- & 0.74 & 0.73& --&0.65 &0.62 &0.62
&0.45$^*$ & 0.65$^*$\\ $\frac{G_A}{G_V}^{\Xi^- \rightarrow \Xi^0}
$ &--&  $-$0.24 & $-$0.27 &--&--&-- &$-$0.25 &$-$0.16 &$-$0.31\\
\hline

$\frac{G_A}{G_V}^{ \Sigma^-\rightarrow n}$ & $-$0.340$\pm$ 0.017 &
$-$0.28 & $-$0.27& 0.38 &0.38& 0.26 &$-$0.25&$-$0.16 &$-$0.31\\
$\frac{G_A}{G_V}^{\Xi^-\rightarrow \Sigma^0}$ &-- & 1.36& 1.16
&--&0.87& 0.91 & 1.26& 0.95&1.27\\
$\frac{G_A}{G_V}^{\Xi^-\rightarrow \Lambda}$ & 0.25 $\pm$0.05 &
0.27 &  0.21&0.21& 0.14&0.32& 0.21& 0.21 &0.21\\
 $\frac{G_A}{G_V}^{\Lambda \rightarrow p}$ &
0.718$\pm$ 0.015  & 0.83& 0.68 & 0.18&$-$0.90&$-$0.94&0.76 & 0.58&
0.74\\
 $\frac{G_A}{G_V}^{\Xi^0 \rightarrow \Sigma^+ }$&
1.21$\pm$0.05 & 1.36&--& 0.38&1.31&1.28& 1.26& 0.95&1.27\\  \hline
 \end{tabular}
 $^*$ Since $f_1$ = 0 for $ \Sigma^{\pm} \rightarrow\Lambda$ in the present case, predictions are given
for $g_1$ values rather than $g_1/f_1$.
 \caption{The axial-vector form factors $G_A/G_V$ in $\chi$CQM$_{{\rm
config}}$ with and without SU(3) symmetry breaking.  } \label{g1f1}
\end{center}}
\end{table}

\begin{table}
\begin{center}
\begin{tabular}{ccccccccc}  \hline
Decay & Data& CVC  & Cabibbo  & CQSM & MSI & $\chi$CQM &
$\chi$CQM$_{{\rm config}}$   & $\chi$CQM$_{{\rm config}}$ with\\
$$ &\cite{PDG}& \cite{renton}& \cite{cabibbo2003} & \cite{tim} &
\cite{yama}& \cite{tommy} & with SU(3) &  SU(3) symmetry\\ &&&&
&&& symmetry & breaking\\ \hline $\frac{f_2}{f_1}^{n\rightarrow
p}$ &-- & 3.71& 1.86 &1.57 & 1.86&3.53&1.70 &2.61 \\ $
\frac{f_2}{f_1}^{\Sigma^- \rightarrow \Sigma^0}$&--& 0.84 &0.53&
0.55& --&1.31 & 0.43& 0.73  \\ $
\frac{f_2}{f_1}^{\Sigma^-\rightarrow \Lambda }$ &--&2.34 & 1.49&
1.24 & 0.81 & 2.73 &1.59$^*$& 2.27$^*$\\ $ \frac{f_2}{f_1}^{\Sigma
^+ \rightarrow \Lambda ^*}$ &--& 2.34& 1.49& 1.24& 0.80 & 2.72  &
1.58$^*$& 2.26$^*$\\ $ \frac{f_2}{f_1}^{\Xi^- \rightarrow \Xi^0} $
&--&$-$2.03& $-$1.43&$-$1.08 & --&$-$2.27&$-$1.64 & $-$2.25\\
\hline $\frac{f_2}{f_1}^{\Sigma^-\rightarrow n}$ &$-$0.97
$\pm$0.14& $-$2.03& $-$1.30 & $-$0.96 & $-$0.88& $-$1.82&$-$1.42&
$-$1.81 \\ $\frac{f_2}{f_1}^{\Xi^-\rightarrow \Sigma^0} $ &--&
3.71& 2.61& 2.02& 1.12& 3.85&1.89 &2.87\\
$\frac{f_2}{f_1}^{\Xi^-\rightarrow \Lambda}$ &--& $-$0.12& 0.09&
$-$0.02 &0.18 & $-$0.06 &$-$0.38 &$-$0.37 \\
$\frac{f_2}{f_1}^{\Lambda \rightarrow p}$ &--& 1.79& 1.07&
0.71&1.07& 1.38  &0.44 & 0.85\\ $\frac{f_2}{f_1}^{\Xi^0
\rightarrow \Sigma^+} $&2.0 $\pm$ 1.3 & 3.71&2.60 &--& 1.35& 3.83
& 1.88 & 2.85\\  \hline
\end{tabular}
$^*$ Since $f_1$ = 0 for $\Sigma^{\pm} \rightarrow\Lambda $ in the
present case, therefore only $f_2$ values are mentioned rather
than $f_2/f_1$. \caption{The weak magnetism form factors
$\frac{f_2}{f_1}$ in $\chi$CQM$_{{\rm config}}$ with and without
SU(3) symmetry breaking.}\label{f2f1}
\end{center}

\end{table}

\end{document}